\magnification=\magstep1

%
\hsize = 6.50truein
\vsize = 8.50truein
\hoffset = 0.0truein
\voffset = 0.0truein
\lineskip = 2pt
\lineskiplimit = 2pt
\overfullrule = 0pt
\tolerance = 2000
\topskip = 0pt
\baselineskip = 18pt
\parindent = 0.4truein
\parskip = 0pt plus1pt
\def\medskip{\vskip6pt plus2pt minus2pt}
\def\bigskip{\vskip12pt plus4pt minus4pt}
\def\smallskip{\vskip3pt plus1pt minus1pt}
\centerline{\bf Ab Initio Evidence for the Formation of Impurity
$\bf d_{3z^2-r^2}$ Holes}
\centerline{\bf in Doped La$_{2-x}$Sr$_x$CuO$_4$}
\bigskip
\centerline{Jason K. Perry,$^{1,2}$ Jamil Tahir-Kheli,$^{1,2}$ and 
William A. Goddard III$^2$}
\centerline{\it $^1$First Principles Research, Inc.$^a$}
\centerline{\it 6327-C SW Capitol Hwy., PMB 250, Portland, OR 97201}
\centerline{\it $^2$Materials and Molecular Simulation Center, Beckman 
Institute}
\centerline{\it California Institute of Technology, Pasadena, CA 91125}
\bigskip
\noindent
{\bf Abstract.}  Using the spin unrestricted
Becke-3-Lee-Yang-Parr density functional, we computed
the electronic structure of explicitly doped La$_{2-x}$Sr$_x$CuO$_4$ ($x$ =
0.125, 0.25, and 0.5).  At each doping level, an impurity hole band is formed
within the undoped insulating gap.  This band is well-localized to 
CuO$_6$ octahedra adjacent to the Sr impurities.
The nature of the impurity hole is $A_{1g}$ in symmetry, 
formed primarily from the $z^2$ orbital on the Cu and $p_z$ orbitals on
the apical O's.  There is a strong triplet coupling of this hole with
the intrinsic $B_{1g}$ Cu $x^2-y^2$/O1 $p_{\sigma}$ hole on the same site.  
Optimization of the $c$ 
coordinate of the apical O's in the doped CuO$_6$ octahedron leads to
an asymmetric anti-Jahn-Teller distortion of the O2 atoms toward the central 
Cu.  In particular, the
O2 atom between the Cu and Sr is displaced 0.26 \AA\ while the O2 atom
between the Cu and La is displaced 0.10 \AA.  Contrary to
expectations, investigation of a 
0.1 \AA\ enhanced Jahn-Teller distortion of this octahedron
does not force formation of an $x^2-y^2$ hole, but instead
leads to migration of the $z^2$ hole to the four other CuO$_6$ 
octahedra surrounding 
the Sr impurity.  This latter observation offers a simple explanation for the
bifurcation of the Sr-O2 distance revealed in x-ray absorption fine structure
data.
\vfill
\eject

\centerline{\bf INTRODUCTION}
\smallskip
The electronic structure of the cuprates obtained
by the local spin density approximation (LSDA)$^1$ holds an unusual position in
the field of superconductivity research.
It is considered a failure because it is unable to
produce the undoped insulating ground state and does not explain the
normal state properties of the doped materials.
Yet despite these significant shortcomings, LSDA has in no small way 
contributed to the consensus view that the only orbitals relevant to
cuprate superconductivity
are the Cu $x^2-y^2$ and O1 planar $p_{\sigma}$ orbitals.  Other orbitals,
in particular the Cu $z^2$ and apical O2 $p_z$ orbitals, were determined
to be well below the Fermi level, and therefore not relevant.  This assumption
colors the interpretation of virtually every experiment and imposes clear 
constraints on theories of superconductivity in these materials.

The endurance of the LSDA results regarding the relevant orbitals near
the Fermi level is odd given that more advanced density functionals
have hinted at a more complicated picture.
In particular, the failure of LSDA regarding the undoped state
was addressed in the early 1990's by several groups.
Svane,$^2$ and later Temmerman, {\it et al.},$^3$ considered a 
self-interaction correction (SIC-LSDA), in which the self-Coulomb term that
a localized electron improperly sees with itself in a conventional LSDA
is removed.  Anisimov {\it et al.}$^4$ 
took a somewhat similar approach with the LSDA+U method, as did Czyzyk
and Sawatzky.$^5$
All achieved some success in producing an insulating gap of the
right magnitude in La$_2$CuO$_4$, correcting the most glaring flaw in the LSDA. 

Perhaps the most important and most overlooked point raised in these studies
is the consistent observation of a marked increase in Cu $z^2$ and O2 $p_z$
character at the top of the valence band.
The basic nature of the undoped hole state (a hybrid
Cu $x^2-y^2$ and O1 $p_{\sigma}$ orbital) is unchanged with respect to the
LSDA.  But these calculations all suggest the formation of at least some
$z^2$ holes upon doping, in direct contradiction to the conventional LSDA
model.

Further evidence for $z^2$ character at the Fermi level appeared when
Anisimov {et al.}$^4$ took their calculations a step farther by considering
the effect of implicit and explicit Sr doping in
La$_{1.75}$Sr$_{0.25}$CuO$_4$.  In both cases the LSDA+U approach found
the doped hole to be a localized Cu $x^2-y^2$/O1 $p_{\sigma}$ hybrid in the
undistorted crystal, in support of the LSDA picture.  However,  when
they allowed a distortion of the apical O2 atoms about the doped 
hole, they found
a local minimum where the hole changed to a localized Cu $z^2$/O2
$p_z$ hybrid.  
This anti-Jahn-Teller state was found to be only 54 meV above the 
global minimum.  
The distortion was constrained to be symmetric (both apical
O2 atoms moved 0.26 \AA\ toward the central Cu) and only performed 
for the implicitly doped case. 
An asymmetric distortion was not investigated nor was a 
distortion in the explicitly doped case. Both of these situations should 
be more favorable for this anti-Jahn-Teller state. 

Recently, an x-ray absorption fine structure (XAFS) 
study by Haskel and Stern, {\it et al.}$^6$ concluded that a majority of
Sr impurities in  La$_{2-x}$Sr$_x$CuO$_4$ induced such an anti-Jahn-Teller 
state,
with the Sr-O2 distance increasing by 0.2 \AA.  A minority of Sr impurites lead
to an enhanced Jahn-Teller distortion, with the Sr-O2 distance decreasing by 
0.1 \AA.  As the results of Anisimov, {\it et al.}$^4$ are somewhat ambiguous, 
this intriguing experimental finding invites a reconsideration of the 
{\it ab initio} description of the doped state.

In a series of papers,$^7$ we have argued against the LSDA single band 
picture and presented a model for the cuprate band structure 
involving the stabilization of the Cu $x^2-y^2$/O1 $p_{\sigma}$ band
(herein referred to as $x^2-y^2$) 
with respect to the Cu $z^2$/O2 $p_z$ band (herein referred to as $z^2$). 
This model leads to a Fermi level crossing of the
narrow $z^2$ band and the broad $x^2-y^2$ band along the (0,0) - ($\pi$,$\pi$)
direction of the 3D Brillouin zone and is consistent with several important
and otherwise anomalous
features of the NMR, ARPES, Hall effect, and Josephson tunneling.

Most recently, we took advantage of some modern developments 
in density functional theory (DFT) 
and looked at the electronic structure of undoped La$_2$CuO$_4$ using
the unrestricted spin form of the Becke-3-Lee-Yang-Parr functional 
(U-B3LYP).$^8$
This hybrid functional, developed since the first LSDA calculations were 
performed
on the cuprates, incorporates both gradient corrections and 20\% Hartree-Fock
exchange.$^9$  It has attained a high level of popularity for
its accurate treatment of molecular systems, and
indeed, this `off-the-shelf' functional successfully 
produced a 2.0 eV insulating band gap in La$_2$CuO$_4$, in excellent
agreement with experiment.  As with the SIC-LSDA
and LSDA+U calculations,$^{2-5}$ a significant increase in Cu $z^2$/O2 $p_z$
character at the top of the valence band was noted.

In this work, we
consider how the U-B3LYP functional describes the doped material,
La$_{2-x}$Sr$_x$CuO$_4$ ($x$ = 0.125, 0.25, and 0.5), using supercells of
size 8$\times$, 4$\times$, and 2$\times$, respectively.  The combination of
this new functional and explicit doping produces a clear {\it ab initio}
picture of the cuprates fundamentally different from LSDA.  At each
doping level, holes are inhomogeneously created on one CuO$_6$ octahedron
adjacent to each Sr impurity.  Localization of the doped holes is consistent
with nuclear quadrupole resonance (NQR) data which clearly resolves two
inequivalent Cu sites in doped La$_{2-x}$Sr$_x$CuO$_4$ and 
La$_2$CuO$_{4+\delta}$.$^{10}$  Our calculations reveal the holes are 
dominated by Cu $z^2$ and O2 
$p_z$ character.   These holes are triplet coupled to the intrinsic holes on 
the doped site, leading to the formation of a spin-polarized impurity band 
within the insulating gap.  Independently optimizing
the $c$ axis coordinates of the O2 atoms, we find an anti-Jahn-Teller distortion
about the doped site similar to the local minimum found by Anisimov, 
{et al.}$^4$.  Furthermore, we find an enhanced Jahn-Teller distortion of 
these same apical O2 atoms 
leads to migration of the $z^2$ hole to different CuO$_6$ octahedra around the 
impurity. 
This latter observation suggests a simple explanation for the bifurcation
of the Sr-O2 distance seen in the XAFS data of Haskel and Stern, {\it et
al}.$^6$
While the important issue of disorder will be
addressed in a separate article, the U-B3LYP picture is unequivocal.
The doped hole in La$_{2-x}$Sr$_x$CuO$_4$ is localized near the impurity
and is $z^2$ in character.

As we have already documented the
superiority of this functional in describing the undoped cuprate, this
new picture must be regarded as a serious challenge to LSDA.  The clear
implication from this work is that models of superconductivity should not
be constrained to consider only the Cu $x^2-y^2$ and O1 $p_{\sigma}$ orbitals.
They must consider the Cu $z^2$ and O2 $p_z$ orbitals as well.

\smallskip
\centerline{\bf RESULTS AND DISCUSSION}
\smallskip

Calculations were performed using a modified version of CRYSTAL98.$^{11}$  To
handle the large number of two electron integrals generated in the $x$ = 0.125
calculations ($>$ 70 gigabytes), we altered the way CRYSTAL98 manages these
files.  We also incorporated Johnson's modification of the
Broyden procedure to improve general 
convergence.$^{12}$  For O, the standard 8-51G Gaussian type basis set was
used.$^{13}$  For Cu, Sr, and La, the Hay and Wadt
effective core potentials (ECP's) were used.$^{14}$  
The valence electrons and outer core (3s and 3p for Cu, 4s and 4p for Sr, 
and 5s and 5p for La) were treated explicitly with these ECP's.
The basis sets used were modified from the original basis
sets of Hay and Wadt, since some functions are too diffuse for
calculations on crystals.  For Cu, the two diffuse S exponents were replaced
by a single exponent optimized to 0.30 in our previous publication on 
undoped La$_2$CuO$_4$.$^8$ The two Cu diffuse P exponents were replaced 
by a single exponent optimized to 0.20.
The basis set was contracted to (3s3p2d) based on atomic Cu(II)
calculations.  For La, the two diffuse S, two diffuse P, and the diffuse
D exponents were removed.  The
basis set was contracted to (2s2p1d) based on atomic La(III) and La(II)
calculations.  To this the core S function of the standard Sr 31G basis 
set$^{13}$ was added, and the combined (3s2p1d) basis set was used for 
both La and Sr sites.  The overall basis set is smaller than that used 
previously,$^8$ but tests
with larger basis sets did not lead to any qualitative change in the results
for either the undoped or $x$ = 0.50 doped materials.

The tetragonal La$_2$CuO$_4$ crystal structure (I4/mmm) was taken from 
Hazen.$^{15}$
The lattice parameters of the conventional cell are $a = b = 3.812$ \AA\ 
and $c = 13.219$ \AA.  All calculations, including undoped, were run in
appropriate supercells ($2\times$ for $x = 0.0$ and $x = 0.5$, $4\times$ for
$x = 0.25$, and $8\times$ for $x = 0.125$).  
The direct lattice vectors for these supercells are given in Table I.  The
positions of the irreducible Cu, O1, and La/Sr atoms in the conventional cell
were taken from Hazen: Cu = (0.0000, 0.0000, 0.0000), O1 =
(0.5000, 0.0000, 0.0000), and La/Sr = (0.0000, 0.0000, 0.3614).  However,
the position
of the O2 atom was optimized for each doping level.  For the undoped
material, the optimal position was calculated to be O2 = (0.0000, 0.0000,
0.1812).  This gives a Cu-O2 distance of 2.40 \AA, in good agreement with
crystallography data of 2.43 \AA.  Optimization of other coordinates was
considered on a limited basis, as noted below.

The undoped band structure, calculated in a $2\times$ supercell (Bmab), is
shown in Figure 1.  Unlike the LSDA, $\uparrow$ and $\downarrow$ spins
localize on adjacent Cu's, producing the expected 2.1 eV band gap.  The
results are in good agreement with SIC-LSDA$^{2,3}$ and LSDA+U$^{4,5}$ band 
structures as 
well as our previous U-B3LYP calculations$^8$ which employed a larger basis 
set.  As described in our previous work, analysis of the
density of states (DOS) indicates the undoped hole is composed almost entirely
from the Cu $x^2-y^2$ and O1 $p_{\sigma}$ orbitals, consistent with the
standard model.  However, there is a notable increase in the Cu $z^2$ and
O2 $p_z$ DOS at the top of the valence band as compared to the LSDA.  The
top 0.15 electrons/unit cell of the valence band are 52\% Cu $z^2$/O2 $p_z$.
This value is slightly larger than what we computed previously (40\%),
perhaps due to the change in basis set or due to the change in the O2
geometry, but the observation of significant $A_{1g}$ orbital character at the 
top of the valence band is in qualitative agreement with other DFT methods
that achieved an insulating ground state.$^{2-5}$
As we've discussed in several publications,$^{7,8}$ this increase in
$z^2$ character at the Fermi level is the result of a stabilization of the
$x^2-y^2$ band due to a proper accounting of on-site Coulomb repulsion.
In the LSDA, it is widely agreed that this Coulomb repulsion is 
overestimated.$^{2-5}$
A consequence of this is that the $x^2-y^2$ band is raised in energy 
relative to the fully occupied bands.  As articulated in our previous work,
this observation alone casts some doubt on the validity of the LSDA single 
band picture.

A common approach to understanding the doped state involves starting with
the band structure for the undoped material and removing the appropriate
number of electrons.
Adopting such a rigid band model with the undoped LSDA band structure would
suggest the system is homogeneously doped by removing additional electrons
from the $x^2-y^2$ band.$^1$
A rigid band model from the undoped U-B3LYP band structure would also
suggest the system is homogeneously doped by forming a mixture of $x^2-y^2$
and $z^2$ holes.  However, the rigid band approach is questionable given
NQR$^{10}$ and XAFS$^6$ data which suggest the doped cuprates are both 
electronically and structurally inhomogeneous. To examine this issue, we have
calculated the U-B3LYP band structures for explicitly doped 
La$_{2-x}$Sr$_x$CuO$_4$.

Doping is achieved by making a substitution of a Sr atom for a single La
atom in the appropriate supercell, breaking the degeneracy of
the Cu sites.  For each doping level ($x =$ 0.125, 0.25,
and 0.5), the same behavior was observed:  doped holes are formed
inhomogeneously in the vicinity of the impurity.
In Figure 2, we show schematically the crystal structure
about the Sr impurity.  There are five Cu atoms surrounding the Sr in a
square pyramidal geometry.  The atom labeled Cu in Figure 2 is at the apex,
and the atoms labeled Cu$'$ and Cu$''$ form the base.  The latter pairs of
Cu atoms are distinguished by spin.
Of these five Cu atoms, it is the apex Cu that is preferentially
doped.  Thus we define the doped unit as the CuO$_6$ octahedron associated 
with this apex Cu.  The holes formed in the doped unit are dominantly 
$A_{1g}$ Cu $z^2$/O2 $p_z$ in character and triplet coupled
to the $B_{1g}$ Cu $x^2-y^2$/O1 $p_{\sigma}$ intrinsic holes on the same
site.  Since there is only one doped site per supercell, symmetry dictates
that all of the doped holes will be of the same spin.  It is important to
realize that this would not be the case in the real system, where the
distribution of doped sites is random.  Indeed, we expect a random distribution
of Sr sites to have a profound effect on the band structure, restoring
approximate single unit cell symmetry in particular.  But there is
no reasonable expectation that this would significantly alter either the
$z^2$ orbital character of the holes or the triplet coupling of these holes
to the $x^2-y^2$ holes.  Understanding these caveats, in these highly
ordered doped materials, the degeneracy of the $\uparrow$ and $\downarrow$
spin bands is lifted.  A new $\downarrow$ spin $z^2$ hole band appears in the
undoped insulating gap, having a width which is dependent on the doping level.
The band structure for $x = 0.5$ is shown in Figure 3.  The minority spin 
($\downarrow$ spin) doped $z^2$ band lies entirely above the Fermi level.

More detailed information on the nature of the doped holes can be obtained
from the DOS given in Figure 4 and Table II.  The nature of
the holes on the undoped sites is relatively unchanged upon doping.  The Cu
$x^2-y^2$ hole character ranges from 0.552 at $x =$ 0.0 to 0.602 at $x =$ 0.50.
A similar small increase in hole character upon doping is also observed
on the neighboring O1 atoms.  In contrast, the nature of the holes on
the doped sites is markedly
different.  There is significant Cu $z^2$ hole character (ranging from
0.341 at $x =$ 0.125 to 0.375 at $x =$ 0.5), as well as O2 $p_z$
hole character (ranging from 0.238 to 0.224 on the O2 between the Cu and Sr
and 0.086 to 0.140 on the O2 between the Cu and La).  Inspection of Figure
4 reveals that this Cu $z^2$/O2 $p_z$ character can be isolated to the
band which appears in the insulating gap upon doping.  This band increases
in width upon doping (from $\approx$0.1 eV at $x =$ 0.125 to $\approx$0.8 eV
at $x =$ 0.5), reflecting an increase in coupling between doped sites
as their concentration increases.  The split in the $x^2-y^2$ $\downarrow$
spin hole band should also be noted, particularly in the $x =$ 0.125 and
$x =$ 0.25 DOS figures.  The lower energy component is dominated by the
doped site $x^2-y^2$ orbital, while the higher energy component is dominated
by the undoped site $x^2-y^2$ orbital.
Interestingly, the Cu $x^2-y^2$ hole character
{\it decreases} on the doped sites (from 0.552 at $x =$ 0.0 to 0.410
at $x =$ 0.125 and 0.425 at $x =$ 0.5).  This decrease in Cu $x^2-y^2$
hole character is compensated by an increase in $p_{\sigma}$ hole character
on the neighboring O1 atoms.  This redistribution of charge in the $x^2-y^2$
band effectively minimizes the formation of unfavorable Cu(III) states
when holes are formed in the $z^2$ band.

As stated earlier, the position of the apical O2 atoms along the
$c$-axis was optimized for
each doping level.  For $x =$ 0.5 there are four unique O2 atoms.  For 
$x =$ 0.25 there are six, and for $x =$ 0.125 there are eight.  The positions
of all four O2 atoms in the $x =$ 0.50 state were optimized independently.  
The O2 atom between the Sr and Cu in the doped unit moved away from
the Sr atom and toward the Cu by 0.24 \AA.  The O2 atom between the La and Cu 
of this doped unit moved toward the Cu by 0.10 \AA.  The O2 atoms of the
undoped CuO$_6$ octahedra both moved a negligible amount in
the direction away from the Sr impurities (0.03 and 0.02 \AA).  Investigation
of Sr and La $c$-axis displacement indicated that the Sr moved toward
the doped CuO$_6$ octahedron by 0.02 \AA\ while the La atoms were unchanged.
Thus, the Sr-O2 distance increased overall by 0.22 \AA\ relative to the La-O2
distance.  Investigation of O1 atom displacements in the $ab$ plane
showed no change in their positions.  The well-known CuO$_6$ tilt was not 
investigated.
Given the small displacements of the undoped O2 atoms and the Sr, we chose to 
only optimize the $c$-axis position of the two O2 atoms in the doped unit for
$x =$ 0.125 and $x = $ 0.25.  The results are virtually identical to the
displacements seen at $x =$ 0.50.  At both $x =$ 0.125 and $x = $ 0.25, 
the O2 atom between the Sr and Cu
moved 0.26 \AA\ toward the Cu, while the O2 atom between the La and Cu moved 
0.11 \AA\ toward the Cu.  While Anisimov, {\it et al.}$^4$ only considered a 
symmetric distortion of the apical O's about the doped site, the 0.26 \AA\ 
displacement that they reported is in agreement with our data.

The finding of a localized hole in these calculations is consistent with
NQR data which reveals two distinct types of Cu's in the doped 
superconductor.$^{10}$  The anomalous 'B' sites have been shown to increase
with doping, and their appearance is independent of the dopant (Sr or
excess O).  The NQR data provide strong support for localized hole formation,
something the standard LSDA band structure fails to predict.$^1$
Analysis by Hammel, {\it et al.}$^{10}$ suggests the hole is
localized in a CuO$_6$ octahedron adjacent to the impurity.
Indeed, a large anti-Jahn-Teller distortion at the apex Cu site around the 
impurity is found in the XAFS data of Haskel and Stern, {et al.}$^6$ 
In some samples, they found an anti-Jahn-Teller distortion of 0.15 \AA\ 
for the O2 atom between the Cu and Sr.$^{16}$  In other samples (differing in
preparation procedure), they found 
a bifurcation of the O2 position, where an anti-Jahn-Teller distortion of
$\approx$0.2 \AA\ was observed at a majority of sites and an enhanced
Jahn-Teller distortion of $\approx$0.1 \AA\ was observed at a minority of
sites.  This structural bifurcation could not be observed in these calculations
due to the symmetry of the supercells employed.  However, 
our calculations are in
quantitative agreement with the anti-Jahn-Teller distortion at the doped site. 

To also gain insight into the enhanced-Jahn-Teller state,
we calculated the band structure for $x =$ 0.125 moving the O2 
atom between the Sr and Cu 0.1 \AA\ toward the Sr.  Given this distortion, we
optimized
the $c$-axis position of the other four O2 atoms about the Sr impurity.
All other atoms were left in their undoped positions.  A comparison of the DOS 
for the anti-Jahn-Teller distorted optimal structure, 
the unoptimized structure, and this enhanced Jahn-Teller 
structure for $x =$ 0.125 is given in Figure 5.  The main change observed in 
the DOS
with the movement of this O2 atom is the shift in the $z^2$ hole band.  For the
anti-Jahn-Teller distorted optimal structure, this $z^2$ band is in the 
middle of the undoped insulating gap.  For the undistorted structure, this 
band is near the bottom of the gap.  But for the enhanced Jahn-Teller 
structure, the $z^2$ band is stabilized such that holes are formed in another 
band.  A detailed analysis of the DOS 
reveals that the $z^2$ holes migrate to either the two Cu$'$ octahedra 
adjacent to the impurity (having a spin opposite to the apex Cu) or
to the two Cu$''$ octahedra adjacent to the impurity 
(having a spin parallel to the apex Cu).  At certain geometries it was
possible to form holes on all five Cu's surrounding the impurity, but
optimization of the O2 atoms led to localization.  Here we have
shown the particular localization where $z^2$ holes form at the Cu$'$ sites.
The apical O's
associated with the Cu$'$ atoms undergo an anti-Jahn-Teller distortion of 
0.09 \AA\ while the apical O's associated with the two Cu$''$ atoms see no
such distortion.  Given the very small gap between the $z^2$ electrons and
the $z^2$ holes on the doped sites, we expect the $z^2$ hole to further
localize to just one of the Cu$'$ octahedra.  Symmetry constraints 
prevent this in these calculations.

With respect to the anti-Jahn-Teller
optimized structure, this enhanced-Jahn-Teller state is 0.57 eV higher
in energy (71 meV per La$_{1.875}$Sr$_{0.125}$CuO$_4$ unit cell).  
While the ground state of this highly ordered doped structure is well defined,
these calculation indicate the doped $z^2$ hole could form at any of
the five Cu atoms surrounding the impurity.
We suggest the observed bifurcation in the Sr-O2 distance
is due to migration of the $z^2$ holes to alternative sites 
when the locations of the impurities is random.  In order to produce
a more uniform distribution of doped holes given random impurities, in some 
cases it may be favorable to form holes at either the Cu$'$ or Cu$''$ sites
leading to an enhanced Jahn-Teller distortion of the apex O2 atom.  
In the majority of cases, holes should be formed at the apex Cu as 
described in this work, leading to an anti-Jahn-Teller distortion of the 
apex O2 atom.  Discrepancies in the XAFS samples among various samples may
be due to doping uniformity, as has already been noted.$^6$
Calculations characterizing the band structure with randomly 
distributed impurities will be documented in an upcoming article.

\smallskip
\centerline{\bf CONCLUSIONS}
\smallskip

In summary, we performed DFT band structure calculations on the
doped superconductor La$_{2-x}$Sr$_x$CuO$_4$ ($x =$ 0.125, 0.25, and 0.5)
using the unrestricted spin
form of the B3LYP functional. In a previous publication,$^8$
this functional was shown to provide a superior description
of the undoped antiferromagnetic state of La$_2$CuO$_4$.
Here, we showed it
leads to a fundamentally different picture of the doped state
as compared to the conventional LSDA description.  Hole formation was found
to be highly inhomogeneous, with localization adjacent to the Sr impurity.
Most significantly, the symmetry of the hole was found to be $z^2$ as opposed
to the commonly believed $x^2-y^2$.  A triplet coupling between the $A_{1g}$
doped holes and the $B_{1g}$ intrinsic holes leads to the formation of a
highly spin polarized $z^2$ band which lies in the middle of the 
insulating gap.  We expect that a random distribution of Sr sites will
profoundly affect the band structure, restoring the spin symmetry and
approximate single unit cell symmetry, but producing significant lifetime
effects.  However, given the robustness of
these results across a wide range of doping levels, we expect the basic nature 
of the doped hole should be largely unchanged from that presented here.

It is our firm view that the LSDA band structures from fourteen years ago
have led the field astray.  At this juncture, the {\it ab initio} evidence
supports the formation of $z^2$ holes upon doping, and this must be taken
into account in the analysis of normal state properties and in the development
of models of superconductivity.

We wish to acknowledge helpful discussions with Dr. Daniel Haskel
and Dr. Peter Schultz.
This work was partially supported by the Materials and
Process Simulation Center (MSC) at Caltech which is supported by grants
from DOE-ASCI, ARO/DURIP, ARO/MURI, 3M, Beckman Institute, Seiko-Epson,
Dow, Avery-Dennison, Kellogg, and Asahi Chemical.
\medskip
\centerline{\bf REFERENCES}
\smallskip
\item{$^a$}http://www.firstprinciples.com.

\item{$^1$}J. Yu, A.J. Freeman, and J.H. Xu, Phys. Rev. Lett. {\bf 58}, 1035 
(1987); L.F. Mattheiss, Phys. Rev. Lett. {\bf 58}, 1028 (1987); W.E. Pickett, 
Rev. Mod. Phys. {\bf 61}, 433 (1989).

\item{$^2$}A. Svane, Phys. Rev. Lett. {\bf 68}, 1900 (1992).

\item{$^3$}W.M. Temmerman, Z.Szotek, and H. Winter, Phys. Rev. B {\bf 47}, 
11533 (1993).

\item{$^4$}V.I. Anisimov,  M.A. Korotin, J. Zaanen, and O.K. Andersen,
Phys. Rev. Lett. {\bf 68}, 345 (1992).

\item{$^5$}M.T. Czyzyk and G.A. Sawatsky, Phys. Rev. B {\bf 49}, 14211 (1994).

\item{$^6$}D. Haskel, E.A. Stern, D.G. Hinks, A.W. Mitchell, J.D. Jorgensen,
Phys Rev. B {\bf 56}, R521 (1997); E.A. Stern, V.Z. Polinger, and D.A. Haskel,
J. Synchrotron Rad. {\bf 6}, 373 (1999); D. Haskel, E.A. Stern, F. Dogan, and
A.R. Moodenbaugh, J. Synchrotron Rad. {\bf 8}, 186 (2001).

\item{$^7$}J.K. Perry and J. Tahir-Kheli, Phys. Rev. B {\bf 58}, 12323
(1998); J. Tahir-Kheli, Phys. Rev. B {\bf 58}, 12307 (1998); J.K. Perry, 
J. Phys. Chem. A {\bf 104}, 2438 (2000); J. Tahir-Kheli, J. Phys. Chem. A 
{\bf 104}, 2432 (2000); J.K. Perry and J. Tahir-Kheli, unpublished 
(cond-mat/9907332); J.K. Perry and J. Tahir-Kheli, J. Phys. Rev B, submitted
(cond-mat/9908308).

\item{$^8$}J.K. Perry, J. Tahir-Kheli, and W.A. Goddard III, Phys. Rev. B
{\bf 63}, 144510 (2001).

\item{$^9$}A.D. Becke, J. Chem. Phys. {\bf 98}, 5648 (1993); C. Lee, W. Yang,
and R.G. Parr, Phys. Rev. B {\bf 37}, 785 (1988).

\item{$^{10}$}P.C. Hammel, B.W. Statt, R.L. Martin, F.C. Chou, D.C. Johnston,
and S.-W. Cheong, Phys. Rev. B {\bf 57}, R712 (1998).

\item{$^{11}$}V.R. Saunders, R. Dovesi, C. Roetti, M. Caus\`a, N.M. Harrison,
R. Orlando, C.M. Zicovich-Wilson, {\it CRYSTAL98 User's Manual}, University
of Torino, Torino, 1998.

\item{$^{12}$}D.D. Johnson, Phys. Rev. B {\bf 38}, 12807 (1988); C.G. Broyden,
Math. Comput. {\bf 19}, 577 (1965).

\item{$^{13}$}http://www.dl.ac.uk/TCS/Software/CRYSTAL.

\item{$^{14}$}P.J. Hay and W.R. Wadt, J. Chem. Phys. {\bf 82}, 299 (1985).

\item{$^{15}$}R.M. Hazen, in {\it Physical Properties of High Temperature
Superconductors II}, ed. D.M. Ginsberg (World Scientific, New Jersey; 1990),
121-198.

\vfil\eject
\noindent{\bf Table I.} Direct lattice vectors for the $2\times$, $8\times$,
$4\times$, and $2\times$ supercells of La$_{2-x}$Sr$_x$CuO$_4$, where
$x = 0.0, 0.125, 0.25,$ and $0.5$, respectively (in \AA).

\vskip 0.5truein
\halign{\noindent#\hfill &\quad \hfill#\hfill &\quad \hfill#\hfill 
&\quad \hfill#\hfill \cr
\noalign{\bigskip\hrule\smallskip}
\noalign{\hrule\medskip}
& -1.906 & 1.906 & 6.609 \cr
$x = 0.0$ & 1.906 & -1.906 & 6.609 \cr
& 3.812 & 3.812 & 0.000 \cr
\noalign{\medskip\hrule\medskip}
& -7.624 & -7.624 & 0.000 \cr
$x = 0.125$ & 7.624 & -7.624 & 0.000 \cr
& 1.906 & 5.718 & 6.609 \cr
\noalign{\medskip\hrule\medskip}
& -7.624 & 0.000 & 0.000 \cr
$x = 0.25$ & 0.000 & 7.624 & 0.000 \cr
& 1.906 & 1.906 & -6.609 \cr
\noalign{\medskip\hrule\medskip}
& -1.906 & 1.906 & 6.609 \cr
$x = 0.50$ & 1.906 & -1.906 & 6.609 \cr
& 3.812 & 3.812 & 0.000 \cr
\noalign{\medskip\hrule\smallskip}
\noalign{\hrule\bigskip}}
\vfill
\eject
\noindent{\bf Table II.} Populations per atom for doped and undoped 
sites.  Doped sites are defined as the CuO$_6$ unit in line with the
Sr impurity along the $c$-axis (see Figure 2.)
Undoped sites refer to all other atoms.

\vskip 0.5truein
\halign{\noindent#\hfill &\quad \hfill#\hfill &\quad \hfill#\hfill 
&\quad \hfill#\hfill &\quad \hfill#\hfill &\quad \hfill#\hfill \cr
\noalign{\bigskip\hrule\smallskip}
\noalign{\hrule\medskip}
doped & Cu $x^2-y^2$ & Cu $z^2$ & O2 $p_z$ & O2' $p_z$ & O1 \cr
$x = 0.125$ & 0.410 & 0.341 & 0.238 & 0.086 & 0.227 \cr
$x = 0.25$ & 0.418 & 0.367 & 0.227 & 0.108 & 0.245 \cr
$x = 0.5$ & 0.425 & 0.375 & 0.224 & 0.140 & 0.267 \cr
\noalign{\medskip\hrule\medskip}
undoped & Cu $x^2-y^2$ & Cu $z^2$ & O2 & O1 \cr
$x = 0.0$ & 0.552 & 0.003 & 0.005 & 0.180 \cr
$x = 0.125$ & 0.567 & 0.005 & 0.008 & 0.195 \cr
$x = 0.25$ & 0.581 & 0.028 & 0.010 & 0.210 \cr
$x = 0.5$ & 0.602 & 0.015 & 0.019 \cr
\noalign{\medskip\hrule\smallskip}
\noalign{\hrule\bigskip}}
\vfill
\eject
\centerline{\bf FIGURE CAPTIONS}
\smallskip
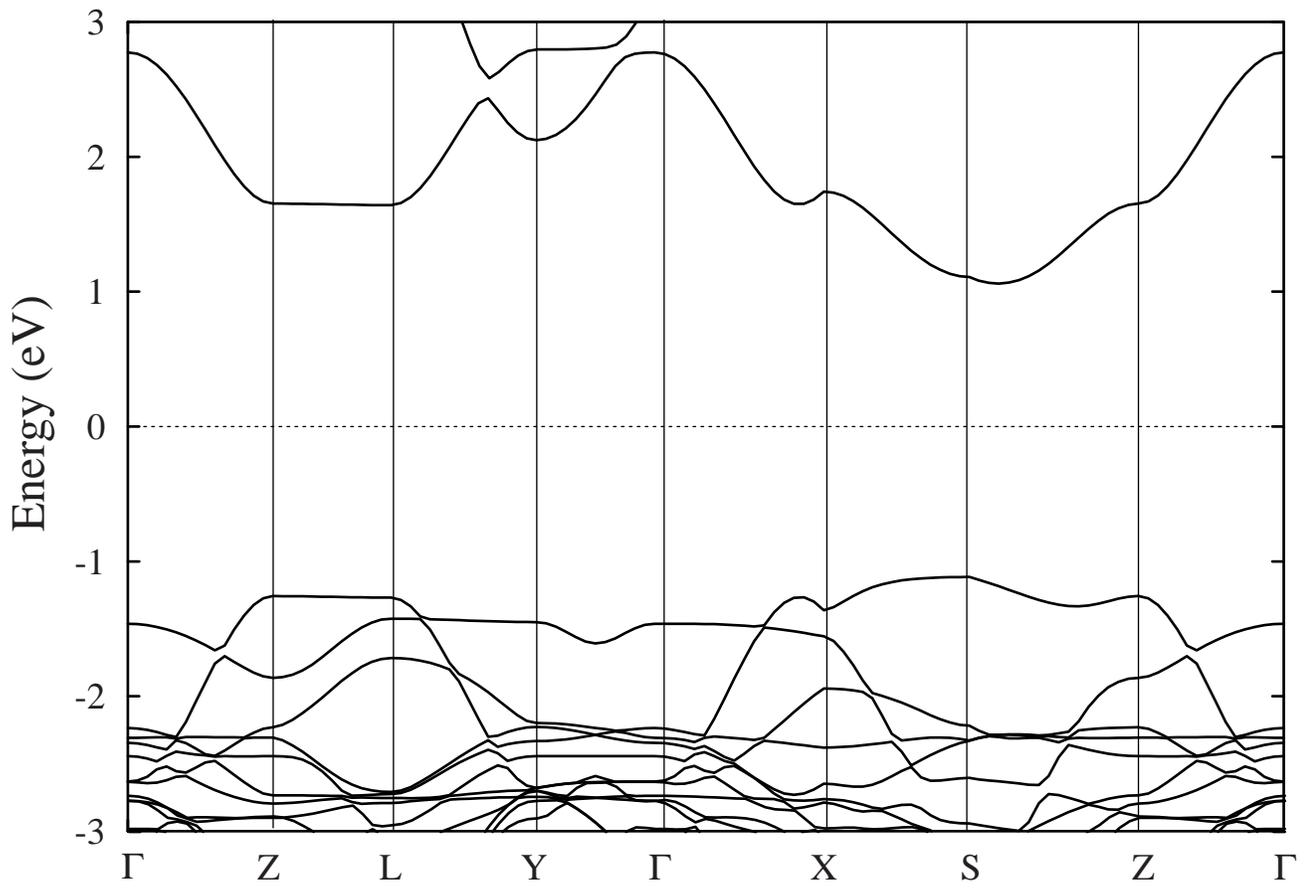
\noindent{\bf Figure 1.}  Band dispersion plotted along symmetry lines of
the orthorhombic Brillouin zone (see reference 8) from the 
U-B3LYP calculation of La$_2$CuO$_4$.  
Results are in good agreement with the SIC-LSD and 
LSDA+U computations of references 2-5.

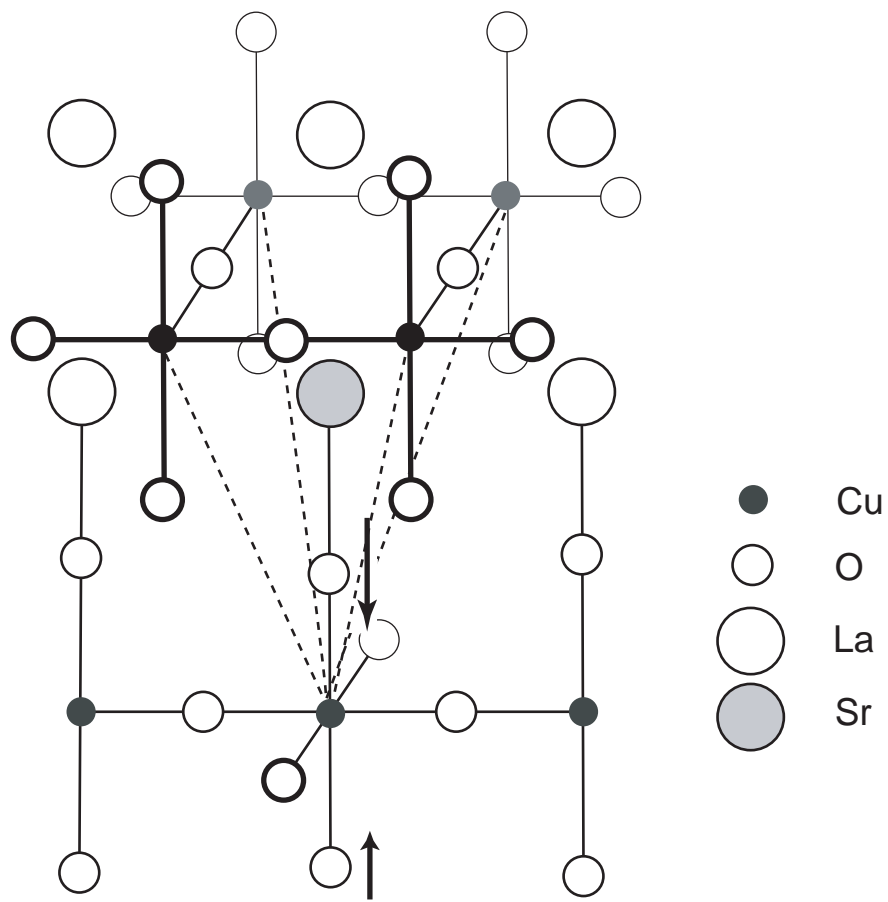
\noindent{\bf Figure 2.}  Schematic of the crystal structure around a Sr 
impurity in La$_{2-x}$Sr$_x$CuO$_4$.  The Cu atom is at the apex of a square
pyramid about the impurity, and Cu$'$ and Cu$''$ form the base.  
The Cu$'$ and Cu$''$ atoms are distiguished by spin.

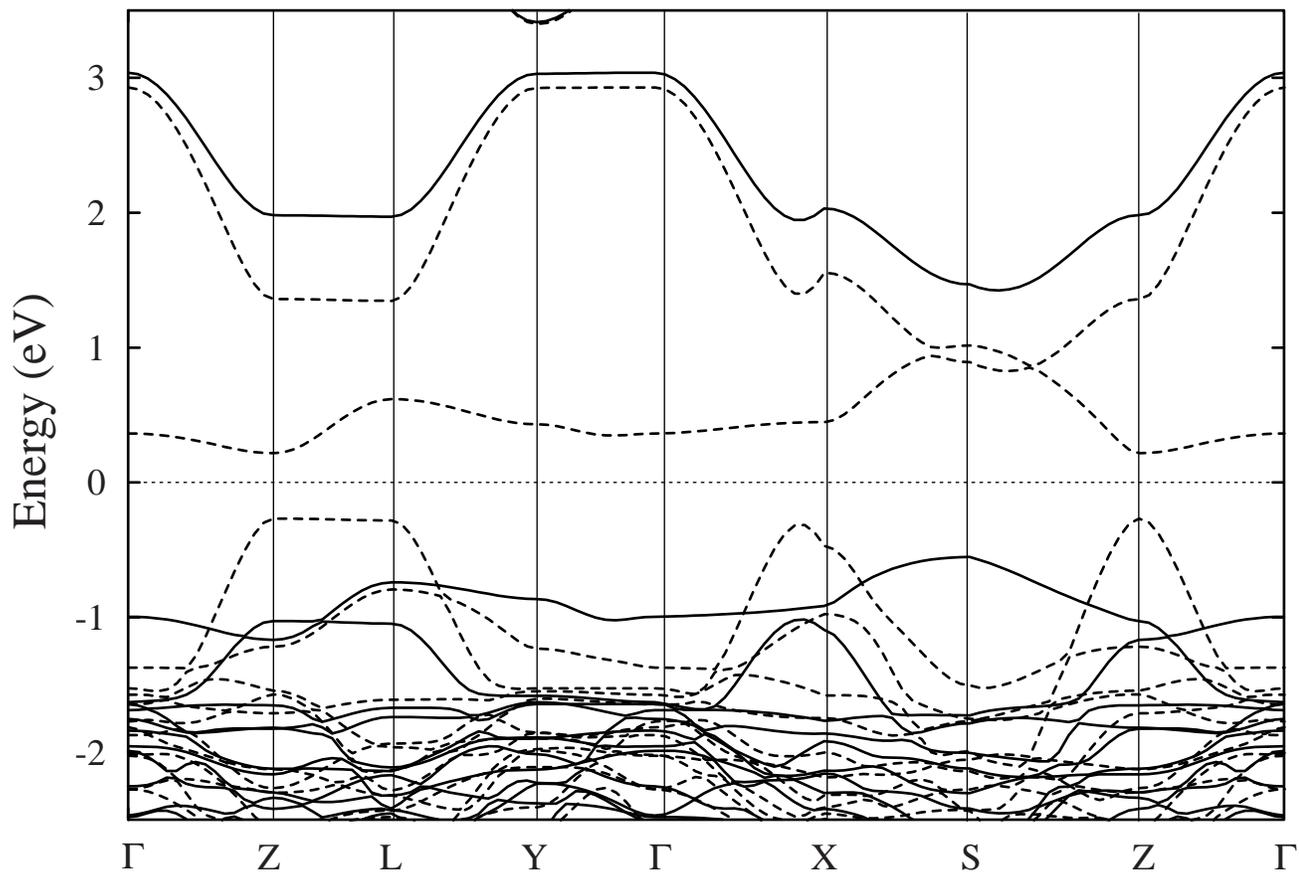
\noindent{\bf Figure 3.}  Band dispersion plotted along symmetry lines of
the orthorhombic Brillouin zone from the 
U-B3LYP calculation of La$_{1.5}$Sr$_{0.5}$CuO$_4$.  Solid line: $\uparrow$
spin.  Dashed line: $\downarrow$ spin.  The splitting of the doped 
$\downarrow$ spin $z^2$ band (the dashed lined just above the Fermi level) 
and the undoped $\uparrow$ spin $z^2$ band (the solid line just below the Fermi
level) is clearly visible.

\noindent{\bf Figure 4.}  Density of states from the U-B3LYP
calculations of La$_{2-x}$Sr$_x$CuO$_4$ (a) $x =$ 0.0, (b) $x =$ 0.125,
(c) $x =$ 0.25, (d) $x =$ 0.5.  Solid line: total DOS.  Dashed line:
Cu $z^2$ $+$ O2 $p_z$ DOS at doped sites.  Dotted line:  Cu $z^2$
$+$ O2 $p_z$ DOS at undoped sites.

\noindent{\bf Figure 5.}  Density of states from calculations on 
La$_{1.875}$Sr$_{0.125}$CuO$_4$ for (a) the anti-Jahn-Teller distorted optimal
structure, (b) the unoptimized (undoped) structure, and (c) the enhanced
Jahn-Teller structure.

\vfill\eject
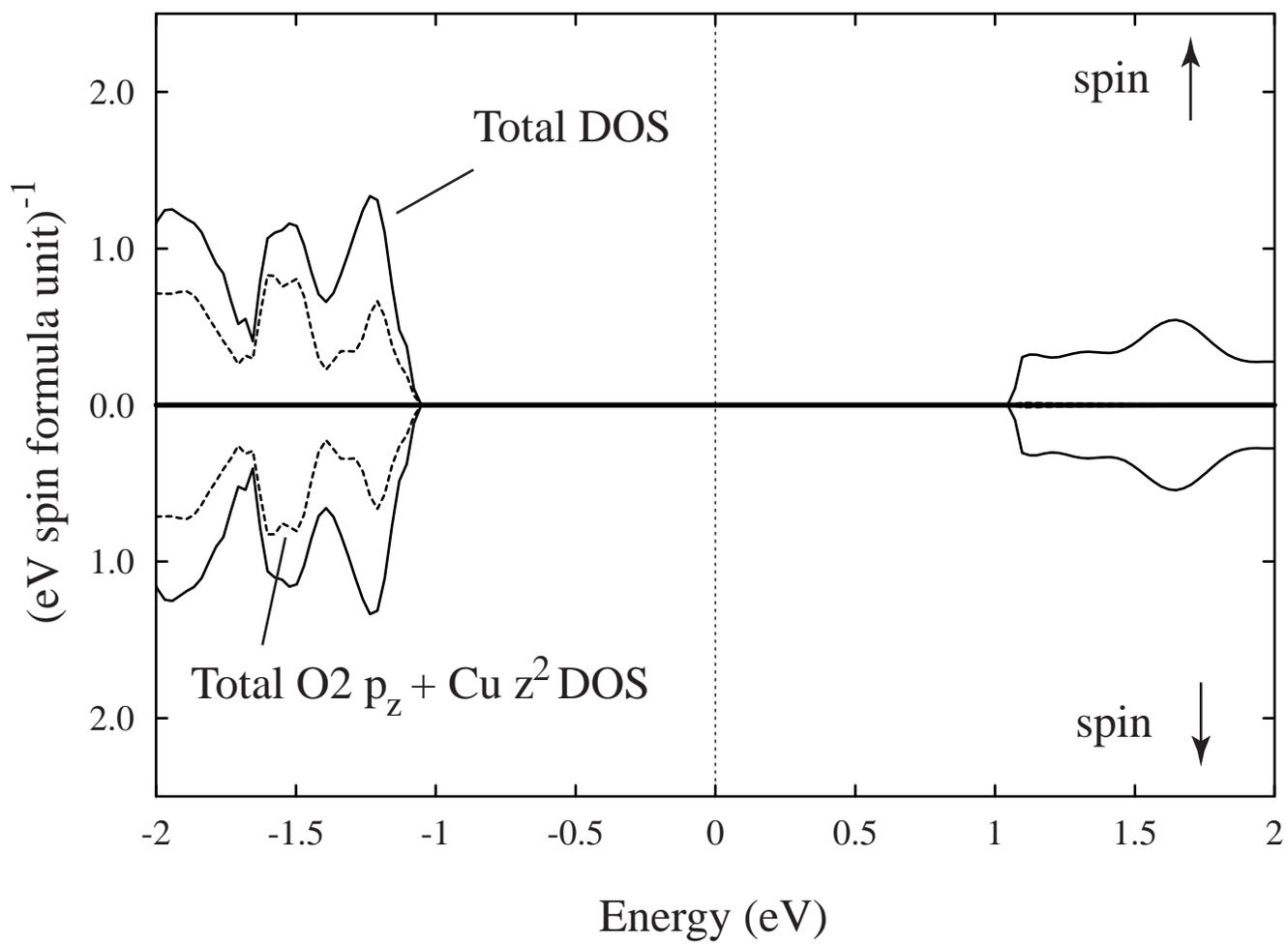
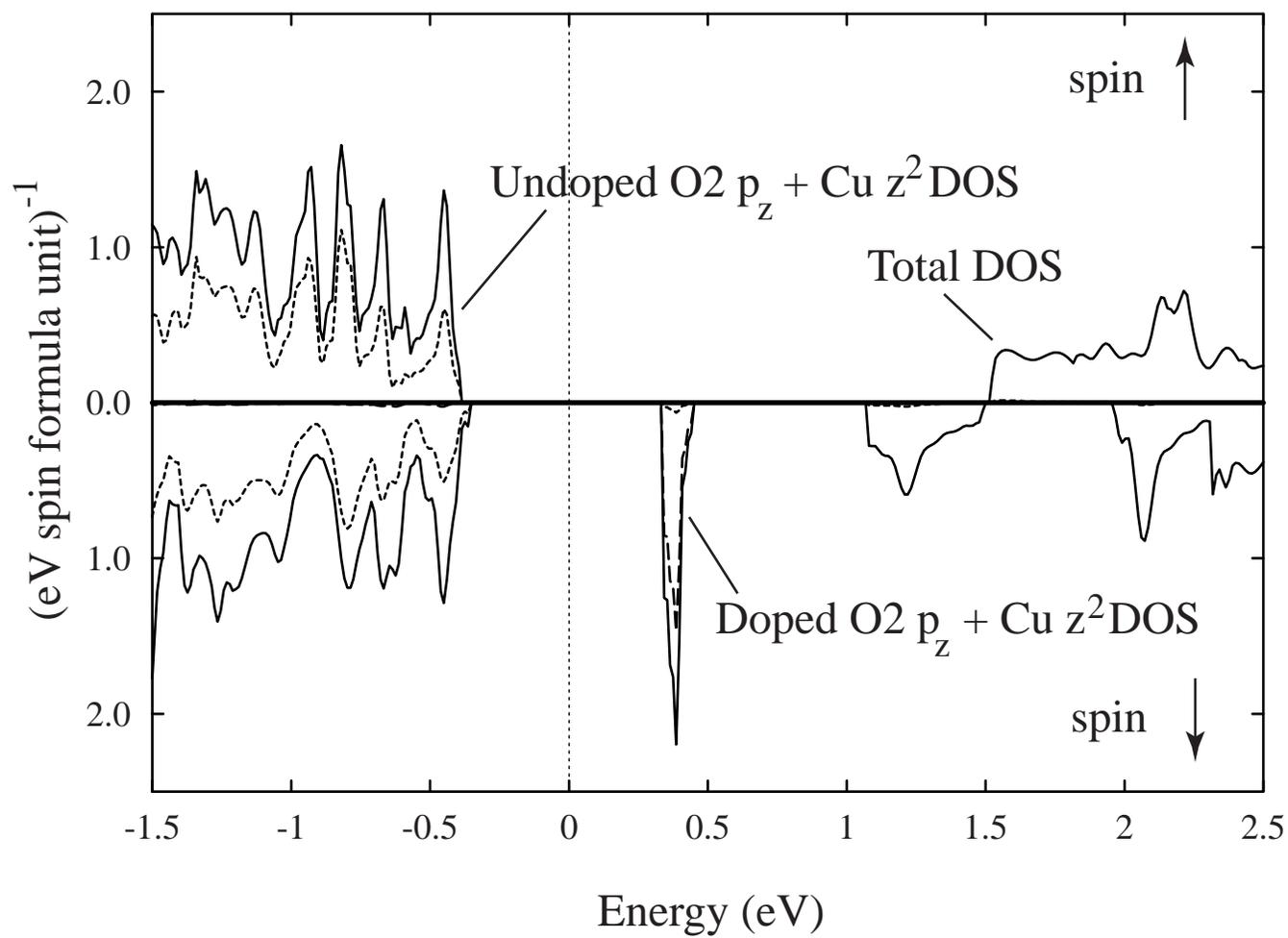
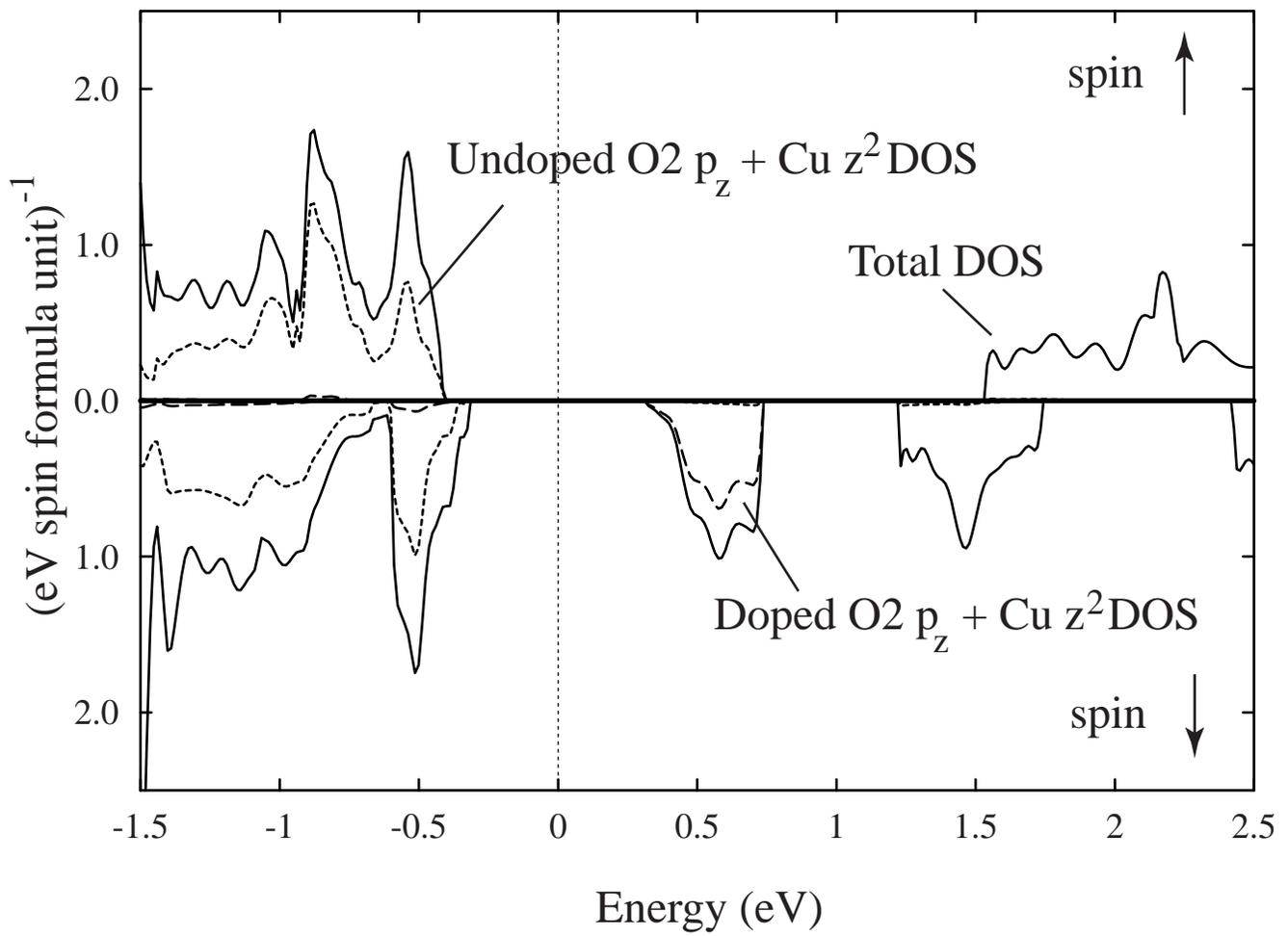
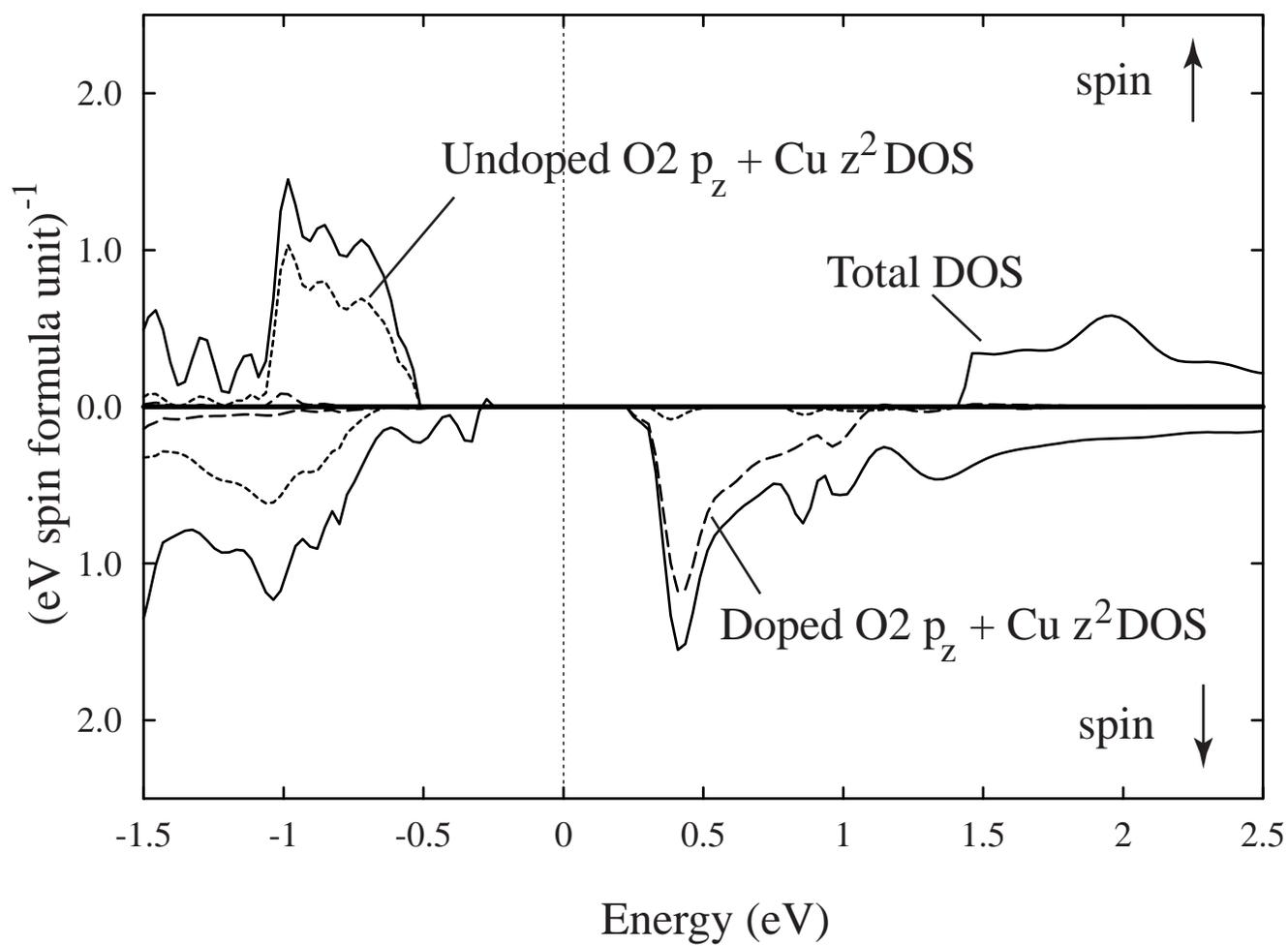
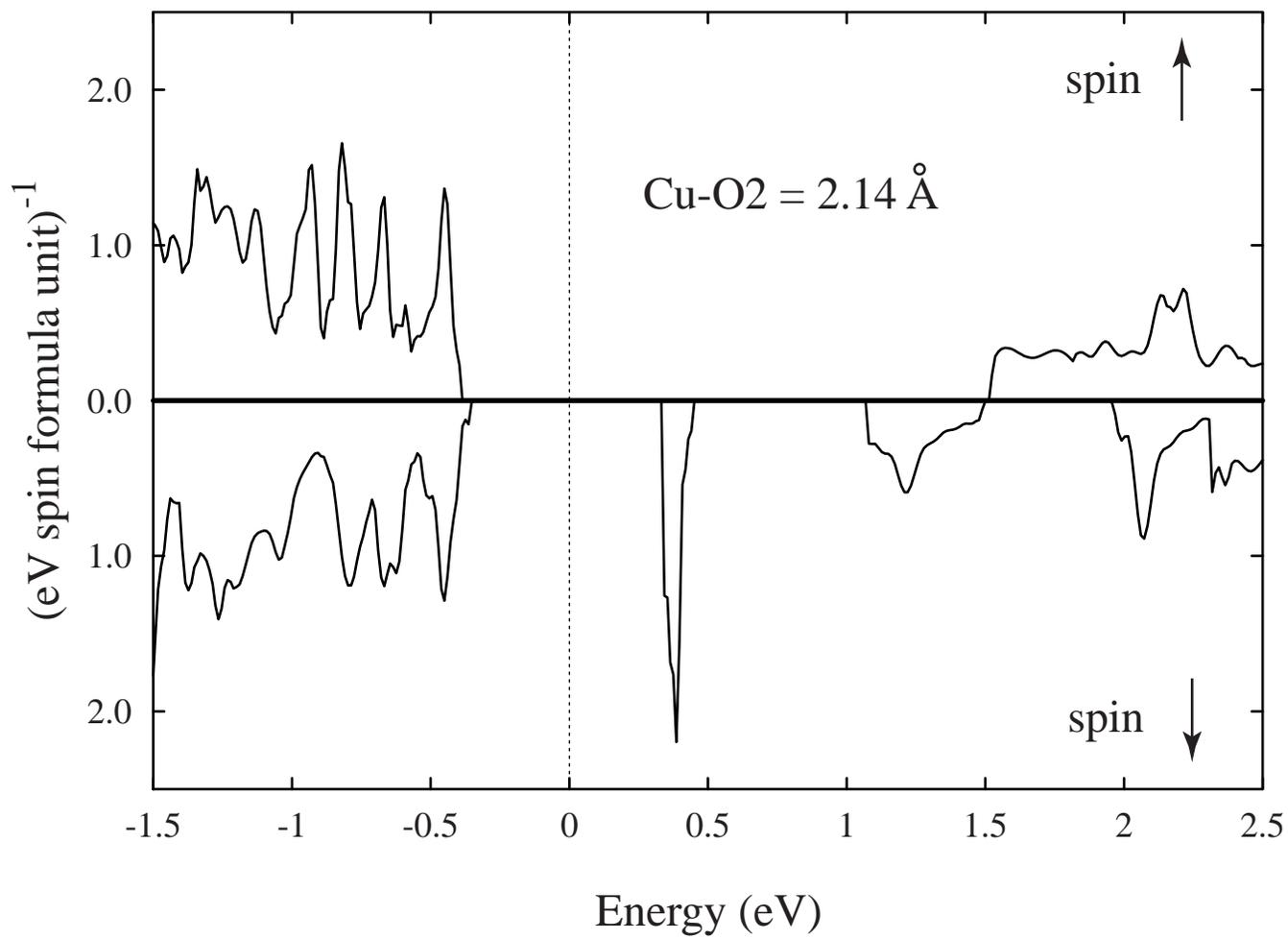
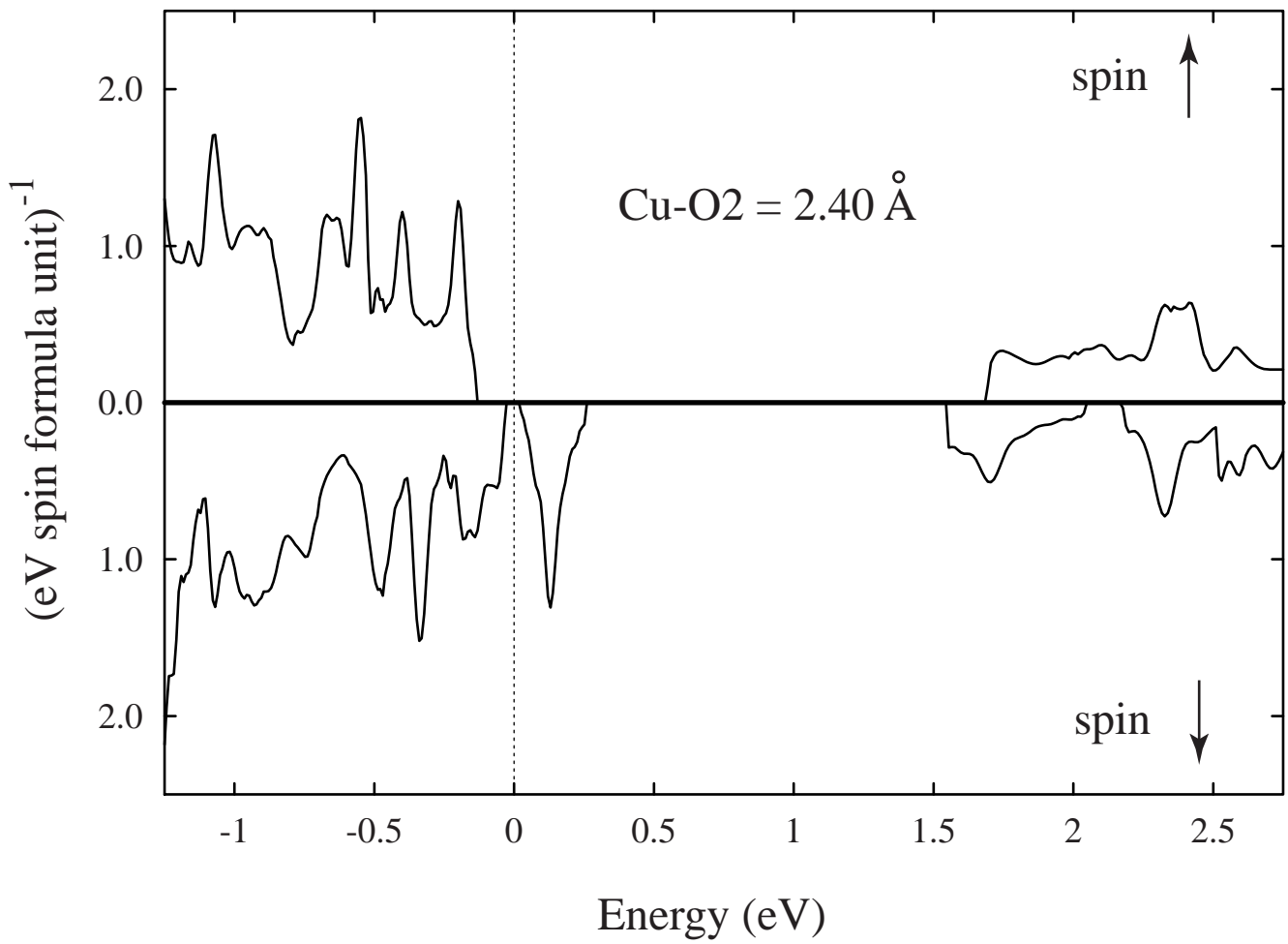
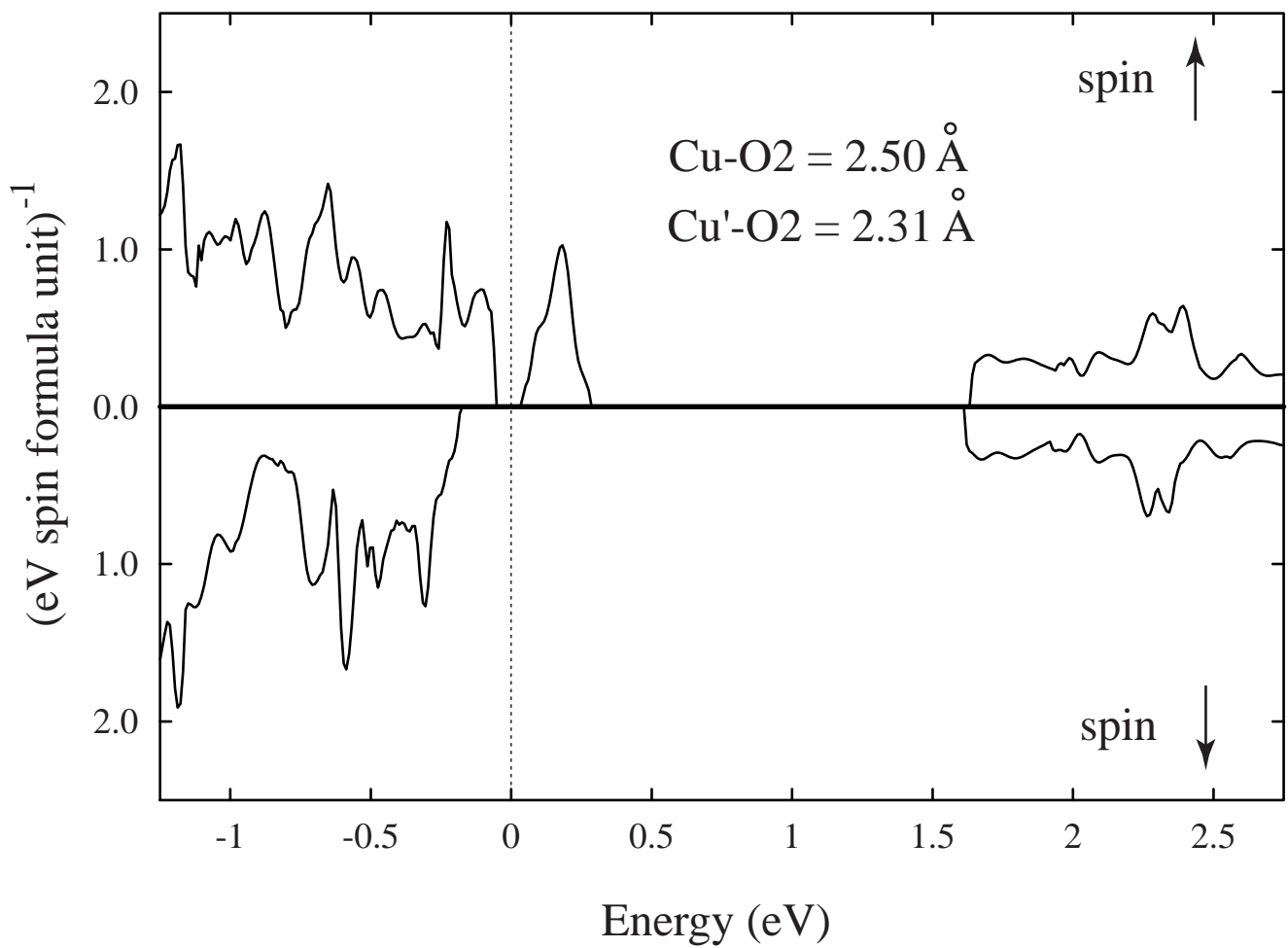
\end